\documentclass{desyproc}
\usepackage{graphicx}
\begin{document}
\title{Beam Test Results with Highly Granular Hadron Calorimeters for the ILC}

\author{{\slshape F. Simon$^{1,2}$, for the CALICE Collaboration}\\[1ex]
$^1$Max-Planck-Institut f\"ur Physik, M\"unchen, Germany\\
$^2$Excellence Cluster ÔUniverseÕ, Garching, Germany}
\contribID{PosterID 44}

\confID{800}  
\desyproc{DESY-PROC-2009-xx}
\acronym{LP09} 
\doi  

\maketitle

\begin{abstract}
To evaluate different technologies for calorimetry at the International Linear Collider, the CALICE collaboration has constructed a highly granular analog hadron calorimeter with small scintillator cells, individually read out by silicon photomultipliers. This device has been extensively tested in particle beams. A digital hadron calorimeter based on RPC readout is currently under construction, with first prototype beam test results already available. The high granularity allows detailed investigations of the substructure of hadronic showers, and can also be exploited for the development of sophisticated reconstruction algorithms.

\end{abstract}

\section{Introduction}

The physics program at a future high energy $e^+e^-$ collider demands excellent reconstruction of multi-jet final states, originating for example from the production and decay of new heavy particles or the hadronic decays of gauge bosons. A promising technique to achieve the necessary jet energy resolution are Particle Flow Algorithms (PFA) \cite{Brient:2002gh, Morgunov:2002pe}, which are based on the reconstruction of individual particles in jets. This approach requires excellent separation of particle showers, and thus extreme granularity, in the calorimeters. Two conceptually different options for PFA-optimized hadron calorimetry are being investigated by the CALICE collaboration, analog and digital sampling calorimetry. The analog option uses small scintillator tiles individually read out with silicon photomultipliers, while the digital option uses gas detectors with small readout pads, which are read out in digital (one bit per channel) or in semi-digital (two or three bits per channel, several amplitude thresholds) mode. Several gas detector technologies are under investigation, such as RPCs and the micro-pattern technologies Micromegas and GEM.

A 1 m$^3$ physics prototype of a scintillator-steel analog hadron calorimeter, with 7608 scintillator cells read out by SiPMs, ranging from $ 3\, \times \, 3$ cm$^2$ to $ 12\, \times \, 12$ cm$^2$ in size, has been tested extensively in particle beams at DESY, CERN and at Fermilab \cite{CALICE:AHCAL}. The imaging capabilities of this calorimeter provide detailed three dimensional information of the hadronic shower activity in the detector. The data recorded with the analog HCAL physics prototype has been used already for a wide range of detailed studies of hadronic showers, with an emphasis on the confrontation of data and simulations using a variety of different hadronic shower models.
A digital hadron calorimeter based on RPC read-out is currently under construction and will be tested in the same absorber structure used for the analog HCAL. First test beam results from small prototypes are already available.

\section{Selected results}

The unprecedentedly granular information available on the topology of the hadronic showers allows the precise measurements of shower profiles, both longitudinally and laterally. 

\begin{wrapfigure}{r}{0.5\textwidth}
\centerline{\includegraphics[width=0.5\textwidth]{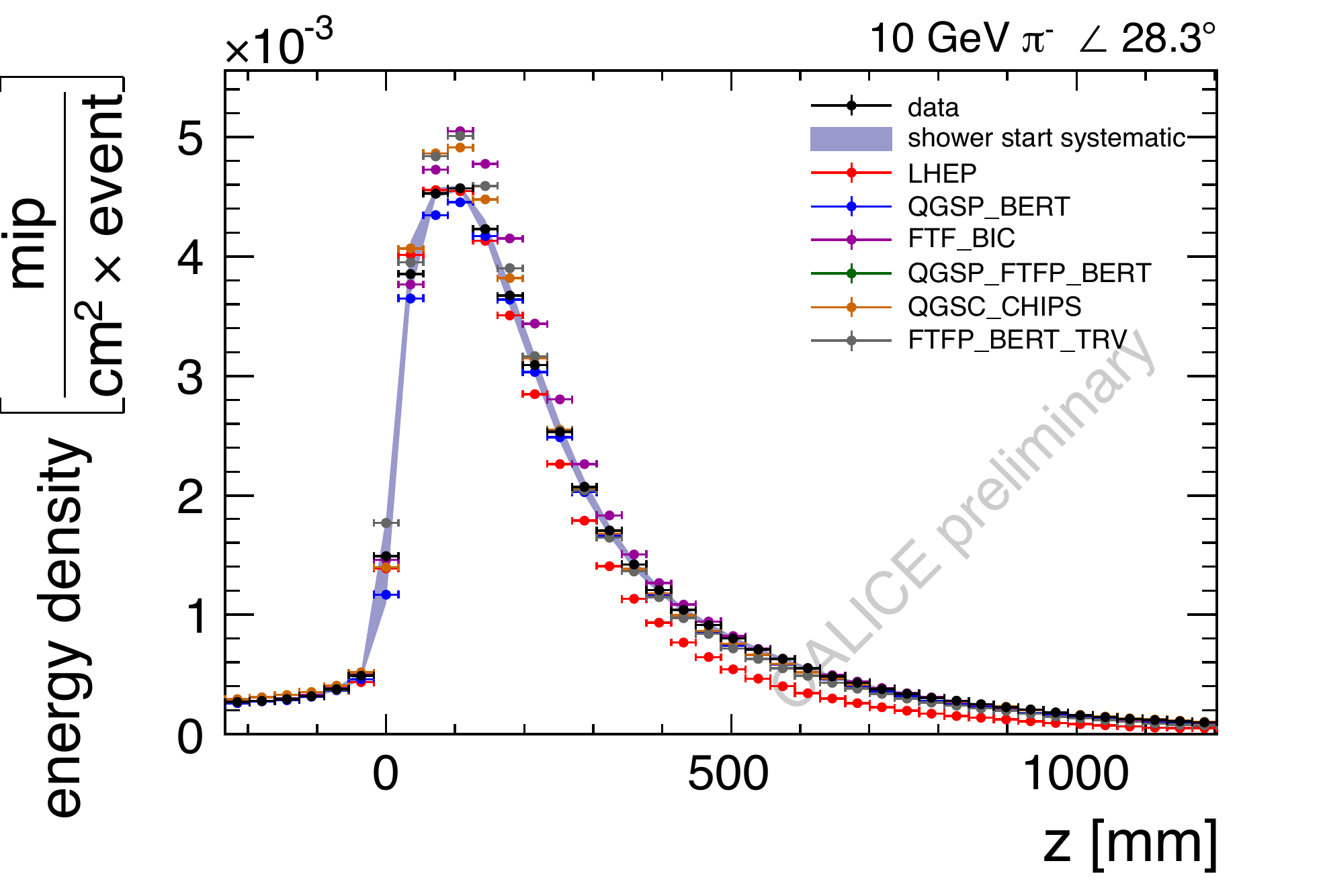}}
\caption{Longitudinal profile of hadronic showers of 10 GeV $\pi^+$, compared to various hadronic shower models.}\label{fig:Profile}
\end{wrapfigure}

Particularly powerful for the comparison of longitudinal shower profiles with simulations is the possibility to directly identify the start point of the shower on an event by event basis. This allows to take the wide distribution of the depth of first interaction in the calorimeter  out of the comparison, significantly improving the sensitivity of the profile studies to differences in the shower models. The position of the shower starting point is identified by increased activity extended over several detector cells, extracted from a specialized clustering algorithm.

Figure \ref{fig:Profile} shows the longitudinal shower profile for 10 GeV $\pi^+$, relative to the start of the shower, measured in the analog HCAL inclined by approximately 30$^\circ$ with respect to the beam axis. The inclination provides an increased depth of the detector, extending the range of the profile measurements. Also shown are simulations using several different physics lists in Geant4 \cite{Geant4Physics}. In particular the theory driven model {\verb}QGSP_BERT}, favored by the LHC test beam campaigns, provides a satisfactory description of the observed profile, while other models show discrepancies in the energy density near the shower maximum and / or in the tails of the distribution. At higher beam energies, significantly increased discrepancies between data and all studied models have been observed, and are under further investigation.

\begin{wrapfigure}{r}{0.5\textwidth}
\vspace{-3mm}
\centerline{\includegraphics[width=0.5\textwidth]{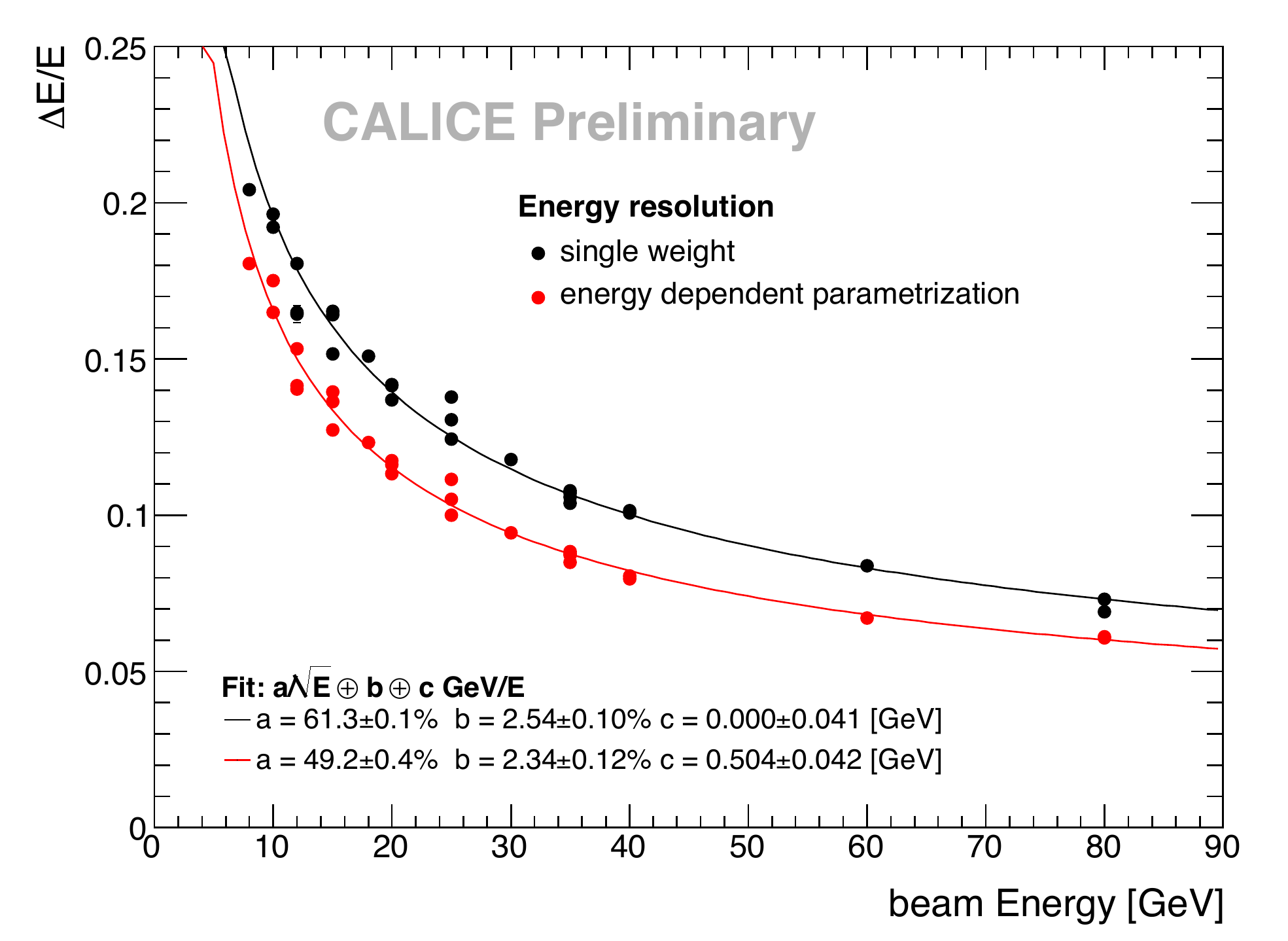}}
\caption{Energy resolution of the CALICE calorimeters for hadrons, with and without software compensation.}\label{fig:Resolution}
\end{wrapfigure}

The high granularity of the analog HCAL is also well suited for the use of software compensation algorithms to improve the hadronic energy resolution, by partially correcting for the difference in detector response to purely hadronic energy deposits and to electromagnetic subshowers. A classification of the different subcomponents can be achieved through the energy density in the shower, since electromagnetic showers tend to be denser than purely hadronic ones. In a first study, a weighting algorithm based on the local shower energy density, given by the amplitude in each individual cell, was \mbox{developed \cite{Simon:2009bt}}. Figure \ref{fig:Resolution} shows the energy resolution for charged pions as a function of beam energy for the complete CALICE calorimeter setup (Silicon-Tungsten ECAL, analog HCAL and tail catcher), both with and without the weighting algorithm. A 20\% improvement in resolution was achieved with the software compensation, which also significantly improves the linearity of the detector response.

On the way to a full physics prototype for a digital HCAL, a small test detector consisting of 6 layers of $20\,\times\,20$ cm$^2$ RPCs with square 1 cm$^2$ readout pads, interleaved with 1.2 $X_0$ absorber plates consisting of 16 mm steel and 4 mm copper, has been tested with muons, electrons and hadrons at Fermilab. The total thickness of the prototype corresponds to 0.65 nuclear interaction lengths, limiting the studies possible with hadrons. 

The electromagnetic data, compared to full detector simulations, demonstrate a good modeling of the detector in Geant4, shown by a satisfactory agreement of measurements and \mbox{data \cite{Bilki:2009ym}}. Some discrepancies, in particular at higher energies in the region of the shower maximum, have been observed and are attributed to loss of efficiency in the RPCs due to high particle densities. Hadronic data has been analyzed in the energy range from 1 GeV to 16 GeV, and the number of observed hits has been compared to detector simulations. Reasonable agreement between data and simulations has been found \cite{Bilki:2009wp}, allowing the extrapolation of the performance of the small prototype to a full calorimeter. These studies demonstrate the potential of this new detector concept, which is now being prepared for experimental verification.

\section{Outlook}

The rich set of hadronic data taken with the physics prototype of a highly granular analog hadron calorimeter for future high energy colliders and the advanced data analysis produce now more and more refined results, demonstrating the capabilities for detailed tests of hadronic shower models as well as moving forward the development of sophisticated reconstruction algorithms. Further refinements of the data analysis and the simulation details promise to provide strong constraints for hadronic shower models within Geant4, and will help in their further development and validation. The first small-scale tests with an RPC based digital HCAL already demonstrated the potential of this new technique, which will be fully explored in the upcoming beam tests of a full physics prototype. This program will provide valuable additional data for the test of shower models, taking into account the insensitivity to neutrons of the used gas detectors, and will be crucial to identify the most promising technology for future experiments.


\begin{footnotesize}


\end{footnotesize}


\end{document}